\documentclass{article}
\usepackage{graphicx}
\usepackage{ai4e}
\usepackage{subcaption}

\usepackage[linesnumbered,ruled,vlined]{algorithm2e}
\usepackage{amsmath}
\usepackage{amssymb}
\usepackage{amsfonts}
\usepackage{fancyhdr}

\usepackage[nameinlink]{cleveref}
\crefname{figure}{Fig.}{Figs.}
\Crefname{figure}{Figure}{Figures}
\crefname{equation}{Eq.}{Eqs.}
\Crefname{equation}{Equation}{Equations}
\crefformat{section}{#2§#1#3}
\crefname{section}{§}{§§}
\Crefname{section}{Section}{Sections}
\crefname{table}{Table}{Tables}
\crefname{appendix}{Appendix}{Appendices}

\newcommand{\eg}{e.g.,\ }
\newcommand{\Eg}{E.g.,\ }

\newcommand{\ie}{i.e.,\ }

\title{Interpretable reinforcement learning for heat pump control through asymmetric differentiable decision trees}

\author{
 Toon Van Puyvelde \\
  IDLab \\
  Ghent University — imec\\
  \texttt{toon.vanpuyvelde@ugent.be} \\
   \And
 Mehran Zareh \\
  Daikin Europe NV\\
  \And
 Chris Develder \\
  IDLab \\
  Ghent University — imec\\
}

\begin{document}
\maketitle
\begin{abstract}
In recent years, deep reinforcement learning (DRL) algorithms have gained traction in home energy management systems. However, their adoption by energy management companies remains limited due to the black-box nature of DRL, which fails to provide transparent decision-making feedback. To address this, explainable reinforcement learning (XRL) techniques have emerged, aiming to make DRL decisions more transparent. Among these, soft differential decision tree (DDT) distillation provides a promising approach due to the clear decision rules they are based on, which can be efficiently computed. However, achieving high performance often requires deep, and completely full, trees, which reduces interpretability. To overcome this, we propose a novel \emph{asymmetric} soft DDT construction method. Unlike traditional soft DDTs, our approach adaptively constructs trees by expanding nodes only when necessary. This improves the efficient use of decision nodes, which require a predetermined depth to construct full symmetric trees, enhancing both interpretability and performance. We demonstrate the potential of asymmetric DDTs to provide transparent, efficient, and high-performing decision-making in home energy management systems.
\end{abstract}


\section{Introduction}
Home Energy Management Systems (HEMs) have gained popularity due to the increasing need for efficient energy usage and cost savings in residential settings. Moreover, from the power grid perspective, to deal with the challenges --- stemming largely from the transition to an increased reliance on renewable sources, as well as growing electrification --- there is a growing interest and need to control residential assets. \Eg dynamic pricing has been introduced to encourage residential energy users to contribute to grid balance through energy flexibility. In recent years, Deep Reinforcement Learning (DRL) algorithms have become prevalent in addressing HEM challenges, due to their ability to optimize complex decision-making processes and adapt to changing environments~\cite{karimi2024jest}.

Despite their potential, adoption of DRL by energy management companies has been limited. This hesitation is primarily due to the black-box nature of DRL algorithms, which fail to provide an interpretable rationale behind their decisions~\cite{hassija2024interpreting}. Such opacity is undesirable for companies that are responsible for their products' actions. In fact, customers may demand insight into the decision-making process of their HEM system. Consequently, explainable reinforcement learning (XRL) techniques have emerged to try and unravel the decision-making processes of traditional DRL algorithms. Other works have approached this problem using post-hoc methods, \eg Local Interpretable Model-agnostic Explanations (LIME)~\cite{kotevska2020methodology} and  SHapely Additive exPlanations (SHAP) values~\cite{heuillet2022collective}.

One promising approach within XRL is the use of decision trees. Tree-based agents determine the next action by traversing a pre-trained decision tree, where nodes branch based on conditions on the state variables~\cite{milani2024explainable,dwivedi2023explainable}. Decision trees offer significant advantages, including linear computation time relative to tree depth and the ability to be validated by experts, ensuring provable safety. Moreover, they provide clear and easily interpretable policies. Using decision trees, we can provide better insight into why an agent selects a particular action, thus enhancing transparency for end users. However, as the depth of decision trees increases, so does their complexity, making them harder to interpret~\cite{izza2020explainingdecisiontrees}. 
To further improve the interpretability of reinforcement learning models, knowledge distillation can be used~\cite{li2020survey,NEURIPS2022_01d78b29}. This method involves training a new interpretable model (the student) to mimic the policy of the original complex DRL model (the teacher). The objective is to retain the teacher model's performance while achieving the student model's interpretability. This approach balances the trade-off between model performance and interpretability, making it a valuable technique for creating transparent and accountable reinforcement learning systems. Using knowledge distillation to distill differentiable decision trees (DDTs) from RL-based agents has been explored in previous works~\cite{frosst2017distillingneuralnetworksoft,soft_ddt_distill, pmlr-v108-silva20a}, and has been introduced in the energy domain by~\cite{gg_ddt}. Despite the potential of distilling soft DDTs from DRL agents, there are notable issues with the current approach:
\begin{enumerate}
    \item The distilled trees are completely full and tend to become harder to interpret, as the number of nodes grows exponentially with tree depth.
    \item The current distillation method occasionally shows instability during training, and this problem becomes more prevalent for deeper trees.
\end{enumerate}
In this work, we propose a solution for these key problems by combining the recursive tree-building approach from the Classification And Regression Tree (CART) algorithm~\cite{breiman1984classification} with the performant knowledge distillation soft DDTs, enabling the construction of \emph{asymmetrical} soft decision trees. Asymmetrical soft decision trees address the first problem by allowing for variable tree depths, which reduces the complexity and improves interpretability. Whereas previous methods consider a fixed tree depth, we use a fixed number of decision nodes. By allowing asymmetric trees, we maximally exploit the model's capacity to mimic the teacher model as closely as possible. In contrast, full trees as created by the existing state-of-the-art method~\cite{soft_ddt_distill} will only reach a limited depth for the same total number of nodes. However, our proposed method allows some nodes to expand deeper than others, aligning with real-world scenarios. Furthermore, by leveraging the strengths of both the CART algorithm and the soft DDT distillation, we intuitively solve the second problem. The recursive tree-building approach from CART provides a robust framework for constructing decision trees. In previous soft DDT distillation methods, all nodes are trained simultaneously, causing changes in high nodes to significantly alter the policy at deeper nodes. In our method, the decision tree is trained node-by-node, providing more stability to the training process as higher nodes remain fixed. This synergy results in more efficient use of decision nodes, making it suitable for practical applications in energy management.
In summary, with this paper we contribute the following:
\begin{itemize}
    \item A novel asymmetric soft DDT construction method that enhances both interpretability and performance in home energy management systems.
    \item Integration of recursive tree-building strategy with the efficient knowledge distillation of soft DDTs.
    \item Variable tree depths, improving the efficient use of decision nodes.
\end{itemize}
\cref{section2} provides an overview of soft differentiable decision trees, including their mathematical formulation and training methodology. This is extended in \cref{section3} with our asymmetric DDT construction method. \cref{section4} formulates the problem environment used to test our approach. Finally, in \cref{section5} we present the experimental results, comparing the performance and explainability of asymmetric DDTs against full trees.

\section{Soft Differentiable Decision Trees} \label{section2}

Soft DDTs are a subclass of machine learning models that combine binary decision trees with differentiability. Unlike traditional decision trees (\eg CART, Random Forests) that make decisions using hard, non-differentiable thresholds, DDTs introduce soft decision boundaries, making them compatible with gradient-descent optimization methods. DDTs have been applied to supervised learning techniques and, more recently, to reinforcement learning (RL) methods~\cite{soft_ddt_distill}.

\subsection{Mathematical Formulation}

Like standard decision trees, a DDT consists of decision nodes and leaf nodes. At the decision nodes, a probabilistic soft function is placed, allowing for continuous-valued routing instead of discrete thresholding. After training, the soft decision nodes are converted into hard nodes by picking the largest weighted feature. At the leaf nodes, a probability distribution is learned over a discrete action space.
\\
\textbf{Soft Routing (Inner Nodes)} DDTs use a sigmoid function:
\begin{equation}
p_\text{L}(\mathbf{x}) = \sigma(\alpha (\mathbf{x} \cdot \mathbf{w}_i + b_i))
\end{equation}
Where, for each node $i$:
\begin{itemize}
    \item $p_\text{L}(\mathbf{x})$ is the probability of taking the left branch,
    \item $\sigma()$ is the sigmoid function,
    \item $\alpha \in \mathbb{R}$ is a temperature parameter controlling sharpness, \ie how distinct the decision boundaries are,
    \item $\mathbf{x} \in \mathbb{R}^d$ is the $d$-dimensional feature vector used for decision-making at the node,
    \item $\mathbf{w}_i \in \mathbb{R}^d$ is the learned weight vector for the feature vector,
    \item $b_i \in \mathbb{R}$ is the learned bias for the node’s decision boundary (split point).
\end{itemize}
The probability of routing to the right branch is given by:
\begin{equation}
p_\text{R}(\mathbf{x}) = 1 - p_\text{L}(\mathbf{x})
\end{equation}
\textbf{Leaf Nodes}
In our DDTs, the leaf nodes output the probability distribution over the discrete action space for the samples reaching that specific node through the decisions of the higher nodes. The action space refers to the set of all possible actions that an agent can take in a given environment. The leaf node distributions are indicated by $Q_\text{L}$ and $Q_\text{R}$. In DDT distillation, the goal is to train these leaf node distributions to closely match the Q-function distribution of the teacher DRL agent.~\footnote{Although in this paper we used DQN, the proposed method is not limited to DQN and it can be used in combination with other RL methods such as SAC and PPO. If the teacher model is an actor-network used in policy gradient methods, leaf node distributions are directly trained to mimic actor network probabilities.}

\subsection{Training Methodology}

Soft DDTs adopt gradient-based learning. The training process involves defining the loss function using the Kullback–Leibler (KL) divergence to measure the discrepancy between the predicted and target distributions, as shown in~\cref{KL_eq}.

\begin{equation}
\label{KL_eq}
    D_{\text{KL}}(P \parallel Q) = \sum_{i \in \mathcal{D}_\text{train}} P(i) \log \frac{P(i)}{Q(i)}
\end{equation}
Gradients of the loss function are then calculated concerning the tree parameters (weights, biases of the decision nodes, and Q values of the leaf nodes) and used to update the model through backpropagation.

\section{Asymmetric Differentiable Decision Trees} \label{section3}

In this work, we introduce a novel training algorithm for constructing asymmetric soft DDTs. This approach aims to use decision nodes more efficiently to approximate a target distribution. Specifically, we iteratively build the tree starting from the root node. Leaf nodes in the current iteration's tree that show a significant difference from the target distribution are further deepened, while those that already closely match the target are not unnecessarily expanded.

\subsection{Recursive Tree Building Algorithm}
Completely full trees, while effective in capturing the policies of deep reinforcement learning models, tend to become increasingly unwieldy as their depth increases. This complexity can make the trees harder to interpret. On the other hand, student models with fewer nodes might not achieve performances as good as that of the teacher model. Asymmetric DDTs can effectively balance this by selectively expanding certain nodes while maintaining some low-depth branches.

Our training algorithm, shown in more detail in~\cref{alg1}, constructs the decision tree iteratively, in a greedy manner. Training starts by training the root node with the full batch of training samples $D_\text{train}$. After that, $D_\text{train}$ is split into $D_\text{L}$ and $D_\text{R}$, containing the training samples that go to respectively the left or right branch by traversing the tree. Consequently, each node is trained using only the samples that would reach that node as filtered by the higher nodes. This isolated training ensures that each node focuses on dividing the state space according to the loss function, based on the relevant samples. In contrast, traditional soft DDTs train all nodes simultaneously, which can lead to inefficiencies as small alterations at upper nodes can cause more drastic changes at deeper nodes. 

Using an iterative distillation method, we limit the depth of the leaf nodes to what is necessary. This approach allows the tree to use a more efficient structure with a constrained number of nodes. It ensures that the decision tree can capture simpler decisions more effectively, reflecting the ease with which the teacher model makes these decisions. As a result, the distilled tree strikes a balance between interpretability and policy accuracy, providing a more transparent and efficient model.

\begin{algorithm}[ht]
\caption{Distilling Asymmetric Differentiable Decision Trees}
\label{alg1}

\KwIn{Training data $\mathcal{D}_\text{train}$, teacher model distribution $Q_\text{teacher}$, max number of decision nodes $N_\text{max}$}
\KwOut{Trained asymmetric differentiable decision tree}

Train root node on $\mathcal{D}_\text{train}$\;
Split $\mathcal{D}_\text{train}$ into $\mathcal{D}_\text{L}$ and $\mathcal{D}_\text{R}$\;

\While{number of decision nodes $< N_\text{max}$}{
    \ForEach{leaf node $\ell$}{
        Calculate $D_\mathrm{KL}\left(Q_\ell \parallel Q_\text{teacher}\right)$\;
    }
    $\ell \gets \arg\max_{\ell'} D_\mathrm{KL}\left(Q_{\ell'} \parallel Q_\text{teacher}\right)$\;
    Train leaf node $\ell$: $\mathcal{D}_\text{train} = \mathcal{D}_\ell$\;
    Split $\mathcal{D}_\text{train}$ into $\mathcal{D}_\text{L}$ and $\mathcal{D}_\text{R}$\;
}
\end{algorithm}

\begin{figure}[ht]
  \centering
  \includegraphics[width=\linewidth, trim=0 0 0 0, clip]{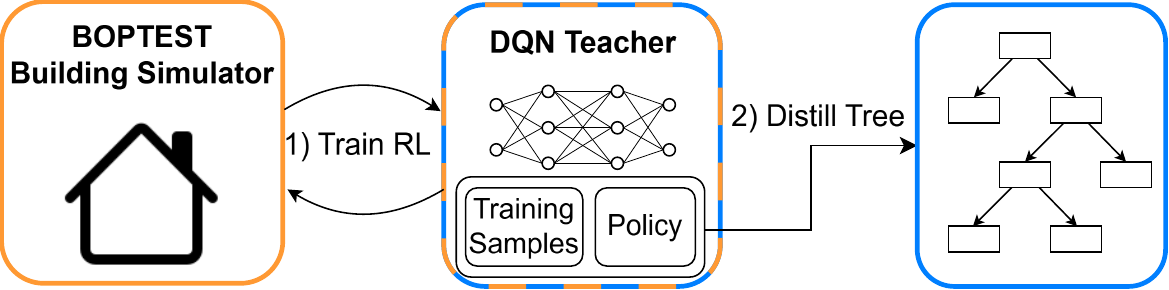}
  \caption{Decision tree distillation overview.}
  \label{fig:architecture}
\end{figure}

\section{Problem formulation} \label{section4}
To test our asymmetric DDT distillation algorithm, we consider a problem environment with sufficient complexity for a tree-based policy to extract more complex intricacies of the teacher policy. We represent a heat pump controller for space heating as a Markov Decision Process (MDP). The state $S_t$ consists of the current hour of the day $h$, zone temperature $T_z$, outside temperature $T_o$, and the indoor temperature's lower comfort bound $T_s$. The action space consists of the heat pump modulation $P_{hp}$ and, as DQN uses a discrete action space, is discretized into 5 elements. The DQN teacher model is trained to maximize the reward function:
\begin{equation}
R_t = -(T_{c,t} + \omega *E_{c,t})
\end{equation}
Where $T_c$ is the increase in thermal discomfort and $E_c$ is the electricity cost KPI of the current time step $t$, and $\omega$ is chosen as 100 to balance operation cost and comfort. As the RL teacher model, we train a DQN agent on the BOPTEST BESTEST hydronic heat pump test case~\cite{boptestgym2021}. Then, from this teacher agent, the asymmetric soft deterministic decision tree distillation is performed as introduced in~\cref{alg1}. We will compare this asymmetric DDT with the full distilled DDT.

\section{Results} \label{section5}
In this section, we compare the results of the experiment for the distilled decision tree against the recursively distilled asymmetric decision tree. Trees with 3, 7, 15, and 31 total decision nodes (corresponding, respectively, to a full tree of depth 2, 3, 4, and 5) are distilled for both trees. The teacher agent and trees are trained on two months of data, and tested on a held-out test set of 14 days. We focus on two main topics. Firstly, comparing the performance of the trees for varying depths. Secondly, we look at interpretability, where we assume a decision tree to be easier to interpret when it contains fewer nodes.

\begin{figure*}[h]
  \centering
  \includegraphics[width=\textwidth]{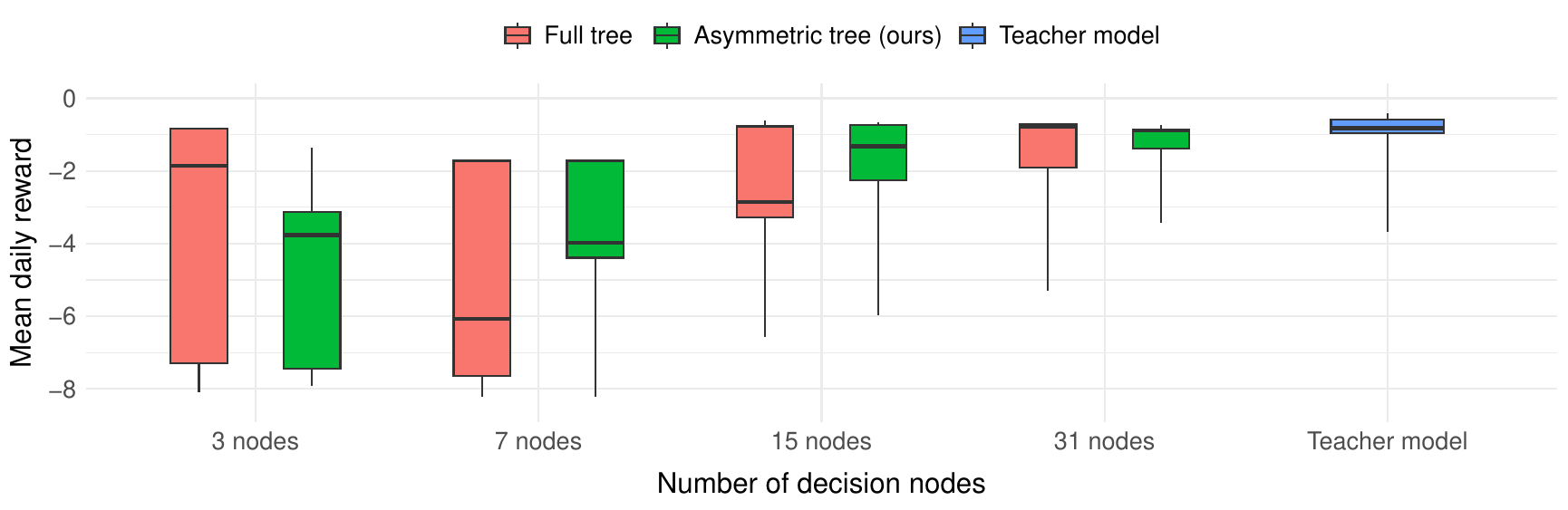}
  \caption{Boxplots comparing the average daily reward over a 2-week test set of completely full DDTs against asymmetric DDTs with the same number of nodes.}
  \label{fig:results}
\end{figure*}

\subsection{Performance of asymmetrical DDT agents}
The results in \cref{fig:results} show that the average rewards from the iteratively built asymmetric DDTs, for trees with sufficient decision nodes, are closer to the rewards of the teacher model than the full distilled trees. For the same number of decision nodes, our approach is thus able to use the available number of nodes more effectively. Because of this, we observe that when increasing the number of nodes, our asymmetric DDTs converge more quickly to a similar performance as the teacher model. 
Additionally, our iterative tree building approach takes away the training instability of full soft DDT training. In full trees, where all nodes are trained simultaneously, changes in upper nodes can significantly alter the tree policy at lower nodes. In our method, the upper nodes are fixed, and thus, sudden instabilities are unable to be caused for the lower nodes.

\subsection{Interpretability of asymmetrical DDT agents}
\begin{figure*}[h]
  \centering
  \includegraphics[width=1\linewidth, trim=20 0 20 0, clip]{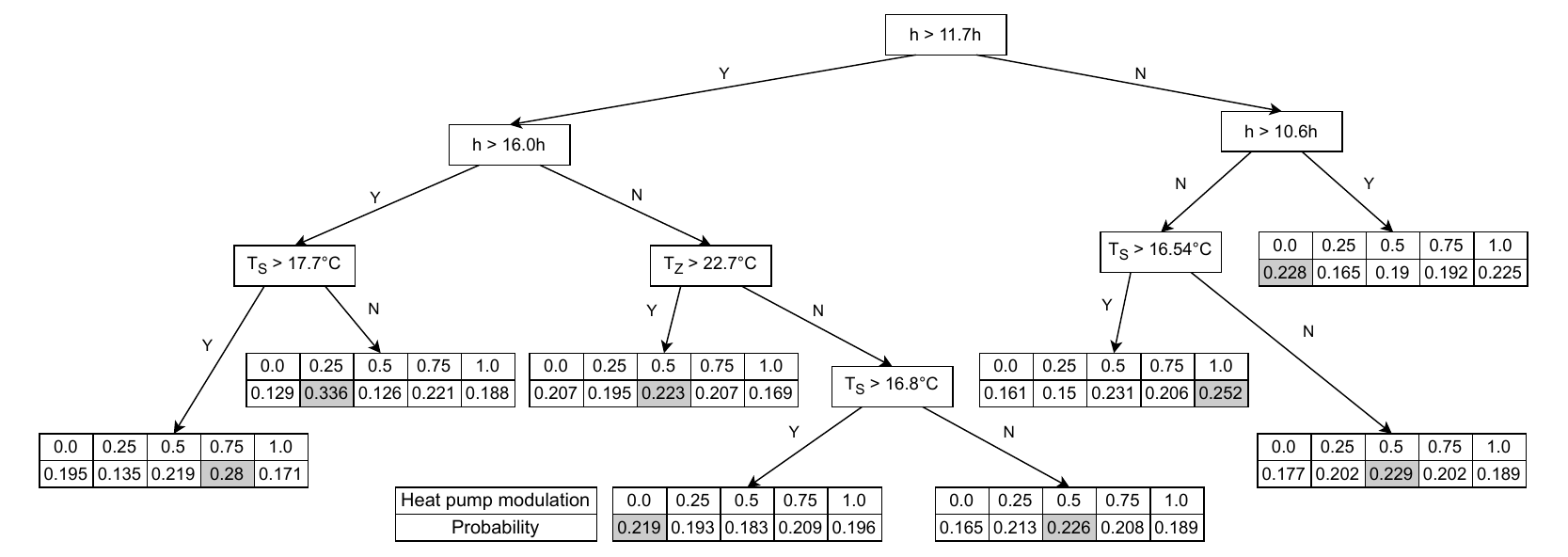}
  \caption{Asymmetric distilled decision tree with 7 decision nodes.}
  \label{tree}
\end{figure*}

Asymmetrical DDTs utilize decision nodes more efficiently by focusing on nodes that differ significantly from the target distribution. This targeted approach ensures that the tree does not become unnecessarily complex, making it easier to interpret by deepening only the leaf nodes that require further refinement. As illustrated in \cref{tree}, the distillation of asymmetric trees gives priority to branches that lie further away from the distribution of the teacher model. This results in trees that perform similarly to full trees, but with considerably fewer decision nodes.

\section{Conclusion} \label{section6}
In this paper, we have explored the integration of asymmetric soft DDTs with DRL to enhance the interpretability and efficiency of decision-making processes in home energy management systems. Our novel training algorithm for constructing asymmetric DDTs demonstrates significant improvements in both interpretability and performance compared to traditional full decision trees.
Our method iteratively builds the decision tree, gradually expanding (deepening) decision nodes that diverge most from the teacher. Thus, our approach ensures that the decision trees remain manageable in size and complexity. This targeted training method allows each node to effectively divide the state space based on relevant samples, enhancing the overall transparency of the model. Our experimental results indicate that asymmetrical DDTs provide interpretable policies, outperforming full distilled trees with the same number of decision nodes. Our iterative approach improves training stability, as each node in the tree is trained without the risk of changes in higher nodes.
Overall, the integration of asymmetric DDTs with DRL offers a promising solution for creating transparent, efficient, and accountable reinforcement learning systems. This approach has the potential to increase the adoption of DRL algorithms by energy management companies, addressing the current limitations posed by the black-box nature of traditional DRL models.
Future work will focus on further refining the training algorithm and exploring its application to other complex environments, aiming to enhance the interpretability and performance of reinforcement learning models across various domains.

\section*{Acknowledgement}
This research was funded by Daikin Europe NV under the PhD Framework Agreement with Ghent University.

\bibliographystyle{unsrt}  
\bibliography{references} 

\begin{thebibliography}{10}

\bibitem{karimi2024jest}
Seyed~Soroush Karimi~Madahi, Bert Claessens, and Chris Develder.
\newblock Distributional reinforcement learning-based energy arbitrage strategies in imbalance settlement mechanism.
\newblock {\em J. Energy Storage}, 104, Part A:114377, Dec. 2024.

\bibitem{hassija2024interpreting}
Vikas Hassija, Vinay Chamola, Atmesh Mahapatra, Abhinandan Singal, Divyansh Goel, Kaizhu Huang, Simone Scardapane, Indro Spinelli, Mufti Mahmud, and Amir Hussain.
\newblock Interpreting black-box models: a review on explainable artificial intelligence.
\newblock {\em Cognitive Computation}, 16(1):45--74, 2024.

\bibitem{kotevska2020methodology}
Olivera Kotevska, Jeffrey Munk, Kuldeep Kurte, Yan Du, Kadir Amasyali, Robert~W Smith, and Helia Zandi.
\newblock Methodology for interpretable reinforcement learning model for hvac energy control.
\newblock In {\em 2020 IEEE International Conference on Big Data (Big Data)}, pages 1555--1564. IEEE, 2020.

\bibitem{heuillet2022collective}
Alexandre Heuillet, Fabien Couthouis, and Natalia D{\'\i}az-Rodr{\'\i}guez.
\newblock Collective explainable ai: Explaining cooperative strategies and agent contribution in multiagent reinforcement learning with shapley values.
\newblock {\em IEEE Computational Intelligence Magazine}, 17(1):59--71, 2022.

\bibitem{milani2024explainable}
Stephanie Milani, Nicholay Topin, Manuela Veloso, and Fei Fang.
\newblock Explainable reinforcement learning: A survey and comparative review.
\newblock {\em ACM Computing Surveys}, 56(7):1--36, 2024.

\bibitem{dwivedi2023explainable}
Rudresh Dwivedi, Devam Dave, Het Naik, Smiti Singhal, Rana Omer, Pankesh Patel, Bin Qian, Zhenyu Wen, Tejal Shah, Graham Morgan, et~al.
\newblock Explainable ai (xai): Core ideas, techniques, and solutions.
\newblock {\em ACM Computing Surveys}, 55(9):1--33, 2023.

\bibitem{izza2020explainingdecisiontrees}
Yacine Izza, Alexey Ignatiev, and Joao Marques-Silva.
\newblock On explaining decision trees.
\newblock {\em arXiv preprint arXiv:2010.11034}, 2020.

\bibitem{li2020survey}
Xiao-Hui Li, Caleb~Chen Cao, Yuhan Shi, Wei Bai, Han Gao, Luyu Qiu, Cong Wang, Yuanyuan Gao, Shenjia Zhang, Xun Xue, et~al.
\newblock A survey of data-driven and knowledge-aware explainable ai.
\newblock {\em IEEE Transactions on Knowledge and Data Engineering}, 34(1):29--49, 2020.

\bibitem{NEURIPS2022_01d78b29}
Wei-Cheng Tseng, Tsun-Hsuan~Johnson Wang, Yen-Chen Lin, and Phillip Isola.
\newblock Offline multi-agent reinforcement learning with knowledge distillation.
\newblock In S.~Koyejo, S.~Mohamed, A.~Agarwal, D.~Belgrave, K.~Cho, and A.~Oh, editors, {\em Advances in Neural Information Processing Systems}, volume~35, pages 226--237. Curran Associates, Inc., 2022.

\bibitem{frosst2017distillingneuralnetworksoft}
Nicholas Frosst and Geoffrey Hinton.
\newblock Distilling a neural network into a soft decision tree.
\newblock {\em arXiv preprint arXiv:1711.09784}, 2017.

\bibitem{soft_ddt_distill}
Youri Coppens, Kyriakos Efthymiadis, Tom Lenaerts, Ann Now{\'e}, Tim Miller, Rosina Weber, and Daniele Magazzeni.
\newblock Distilling deep reinforcement learning policies in soft decision trees.
\newblock In {\em Proceedings of the IJCAI 2019 workshop on explainable artificial intelligence}, pages 1--6, 2019.

\bibitem{pmlr-v108-silva20a}
Andrew Silva, Matthew Gombolay, Taylor Killian, Ivan Jimenez, and Sung-Hyun Son.
\newblock Optimization methods for interpretable differentiable decision trees applied to reinforcement learning.
\newblock In {\em International conference on artificial intelligence and statistics}, pages 1855--1865. PMLR, 2020.

\bibitem{gg_ddt}
Gargya Gokhale, Seyed~Soroush Karimi~Madahi, Bert Claessens, and Chris Develder.
\newblock Distill2explain: Differentiable decision trees for explainable reinforcement learning in energy application controllers.
\newblock In {\em Proceedings of the 15th ACM International Conference on Future and Sustainable Energy Systems}, pages 55--64, 2024.

\bibitem{breiman1984classification}
Leo Breiman, Jerome Friedman, Richard~A Olshen, and Charles~J Stone.
\newblock {\em Classification and regression trees}.
\newblock Routledge, 2017.

\bibitem{boptestgym2021}
Javier Arroyo, Carlo Manna, Fred Spiessens, Lieve Helsen, D~Saelens, J~Laverge, W~Boydens, and L~Helsen.
\newblock An openai-gym environment for the building optimization testing (boptest) framework.
\newblock In {\em Proceedings of Building Simulation 2021: 17th Conference of IBPSA}, volume~17, pages 175--182. INT BUILDING PERFORMANCE SIMULATION ASSOC-IBPSA, 2022.

\end{thebibliography}
\end{document}